\documentclass[%
 reprint,
superscriptaddress,
 amsmath,amssymb,
 aps,
 pra,
]{revtex4-2}

\usepackage{graphicx}

\usepackage{pgfplots}
\pgfplotsset{compat=1.18}

\usepackage{dcolumn}
\usepackage{bm}
\usepackage{hyperref}
\hypersetup{
	colorlinks,
	linkcolor={black},
	citecolor={black},
	urlcolor={blue}
}
\usepackage{tikz}
\usetikzlibrary{decorations.pathreplacing}
\usetikzlibrary{arrows.meta}
\usetikzlibrary{decorations.pathmorphing}
\usetikzlibrary{positioning}
\usetikzlibrary{calc}
\usetikzlibrary{patterns}

\usepackage{adjustbox}
\usepackage{makecell}

\usepackage{xfrac}

\usepackage{booktabs}

\usepackage[T1]{fontenc}   
\usepackage{lmodern}        

\newcommand{\Evec}{\bm{\mathcal{E}}}

\newcommand{\V}{\,\mathrm{V}}

\newcommand{\Vcm}{\,\mathrm{V/cm}}
\newcommand{\kV}{\,\mathrm{kV}}
\newcommand{\cm}{\,\mathrm{cm}}

\newcommand{\kHz}{\,\mathrm{kHz}}
\newcommand{\MHz}{\,\mathrm{MHz}}
\newcommand{\GHz}{\,\mathrm{GHz}}

\newcommand{\mm}{\,\mathrm{mm}}
\newcommand{\nm}{\,\mathrm{nm}}
\newcommand{\mW}{\,\mathrm{mW}}
\newcommand{\W}{\,\mathrm{W}}

\newcommand{\kelvin}{\,\mathrm{K}}

\newcommand{\vms}{\,\mathrm{m/s}}

\newcommand{\CENTREX}{CeNTREX}

\newcommand{\APone}{\textsc{AP1}}
\newcommand{\APtwo}{\textsc{AP2}}

\newcommand{\RtwoFfour}{$R(2)\,\tilde{F}^\prime_1=7/2\,F^\prime=4$}

\newcommand{\PtwoFone}{$P(2)\,\tilde{F}^\prime_1=3/2\,F^\prime=1$}
\newcommand{\PtwoFtwo}{$P(2)\,\tilde{F}^\prime_1=3/2\,F^\prime=2$}

\newcommand{\spinup}{\ket{\uparrow}}
\newcommand{\spindown}{\ket{\downarrow}}

\newcommand{\gnd}{X\,^1\Sigma^+_0}

\renewcommand{\vec}[1]{\mathbf{#1}}

\newcommand{\ket}[1]{\left|#1\right\rangle}

\newcommand{\epsone}{\epsilon_{01}}
\newcommand{\epstwo}{\epsilon_{12}}
\newcommand{\epstot}{\epsilon_{02}}

\newcommand{\rottemp}{$6.3(2)\kelvin$}

\newcommand{\epsonedep}{0.95(1)}
\newcommand{\epsoneacc}{0.88(11)}
\newcommand{\epsoneweighted}{0.92(6)}
\newcommand{\epstwodep}{0.99(1)}
\newcommand{\epstwoacc}{1.11(10)}
\newcommand{\epstwoweighted}{1.05(5)}
\newcommand{\epstotalacc}{0.97(8)}

\newcommand{\nphotonRzeroFtwo}{1.92}
\newcommand{\nphotonRoneFthree}{2.10}
\newcommand{\nphotonRtwoFfour}{2.13}

\usepackage{xspace}

\newcommand{\acc}{app\xspace}

\newcommand{\accumulation}{appearance\xspace}

\newcommand{\dep}{rem\xspace}

\newcommand{\depletion}{removal\xspace}

\begin{document}

\preprint{APS/PRA}

\title{Adiabatic passage of \texorpdfstring{$^{205}$TlF}{205TlF} with microwaves in a cryogenic beam}
\author{Olivier Grasdijk}
\email{jgrasdijk@argonne.gov}
\affiliation{Physics Division, Argonne National Laboratory, Argonne, Illinois 60439, USA}

\author{Jakob Kastelic}
\affiliation{Department of Physics, Yale University, New Haven, Connecticut 06511, USA}

\author{Jianhui Li}
\affiliation{Department of Physics, Columbia University, New York, New York 10027, USA}

\author{Oskari Timgren}
\affiliation{Department of Physics, Yale University, New Haven, Connecticut 06511, USA}

\author{Konrad Wenz}
\affiliation{Department of Physics, Columbia University, New York, New York 10027, USA}

\author{Yuanhang Yang}
\affiliation{Department of Physics, University of Chicago, Chicago, Illinois 60637, USA}

\author{Perry Zhou}
\affiliation{Department of Physics, Columbia University, New York, New York 10027, USA}

\author{David Kawall}
\affiliation{Department of Physics, University of Massachusetts Amherst, Amherst, Massachusetts 01003, USA}

\author{Tanya Zelevinsky}
\affiliation{Department of Physics, Columbia University, New York, New York 10027, USA}

\author{David DeMille}
\email{david.demille@jhu.edu}
\altaffiliation{Also at Physics Division, Argonne National Laboratory, Argonne, Illinois 60439, USA and Department of Physics, University of Chicago, Chicago, Illinois 60637, USA}
\affiliation{Department of Physics and Astronomy, Johns Hopkins University, Baltimore, Maryland 21218, USA}

\collaboration{CeNTREX Collaboration}
\noaffiliation

\date{\today}

\begin{abstract}
We present a hyperfine-resolved state preparation scheme for thallium fluoride (TlF) molecules based on microwave-driven adiabatic passage (AP) in a spatially varying electric field. This method enables efficient and robust population transfer between selected $\ket{J,m_J=0}$ hyperfine sublevels of the $\gnd$ ground state in a cryogenic molecular beam, a key requirement for the CeNTREX search for nuclear time-reversal symmetry violation. Two sequential stages of AP are implemented. The first transfers population from $J=0$ to $J=1$ at a local field of $173~\Vcm$, and the second transfers from $J=1$ to $J=2$ at $110~\Vcm$. Transfer efficiencies are quantified through laser-induced fluorescence, and accounting for residual population in excited rotational levels after a prior stage of rotational cooling. We achieve state transfer efficiencies of $\epsoneweighted$ and $\epstwoweighted$ for the first and second states of AP, respectively. This corresponds to a total efficiency of $\epstotalacc$ for population transfer from $J=0$ to $J=2$. These results demonstrate robust and high-fidelity preparation of specific rotational/hyperfine states in TlF.

\end{abstract}


\maketitle


\section{Introduction}

\CENTREX\ (the Cold Molecule Nuclear Time-Reversal Experiment) seeks to achieve a significant improvement in sensitivity beyond the current best upper limits on the strength of certain hadronic time-reversal ($T$)-violating fundamental interactions, such as the proton electric dipole moment and the $CP$-violating quantum chromodynamics parameter, $\bar{\theta}_{\rm QCD}$. The overall \CENTREX\ approach and measurement strategy are discussed in~\cite{Grasdijk_2021}. By conducting magnetic resonance measurements on the $^{205}$Tl nucleus within an electrically polarized thallium monofluoride (TlF) molecule, \CENTREX\ aims to determine the Schiff moment of $^{205}$Tl. The strongly polarized electron shells in TlF couple strongly to the Schiff moment, resulting in magnetic resonance frequency shifts that are orders of magnitude larger than those achievable in atomic experiments, for the same size of Schiff moment~\cite{sandars1967measurability}.

The \CENTREX\ protocol requires preparing TlF molecules in a variety of different rotational and hyperfine states at different stages of the experiment: efficient rotational cooling~\cite{vb31-f2c3}, molecular beam collimation via electrostatic lenses~\cite{Grasdijk_2021}, implementation of the Schiff-moment measurement sequence~\cite{Hinds1980ExperimentFluoride}, and nuclear-spin readout via optical cycling~\cite{Grasdijk_2021}. CeNTREX plans to implement transfer between these states using microwave-driven adiabatic passage (AP).

Adiabatic passage 
is a robust method for coherently transferring population between quantum states~\cite{treacy_adiabatic_1968, shore_2009, loy_observation_1974}. Unlike $\pi$-pulses, which require precise control of intensity and timing, AP achieves near-unity transfer efficiency whenever the adiabatic condition is satisfied, making it ideal for molecular beam experiments with spatial and velocity inhomogeneities~\cite{budker2004atomic}. Here we demonstrate a state-transfer scheme for \CENTREX\ that employs microwave AP in spatially varying electric fields to transfer population from specific hyperfine sublevels of $\ket{J=0}$ to other specific hyperfine states in the $\ket{J=2}$ manifold of states, all within the TlF ground electronic state.

Adiabatic passage is closely related to the Landau–Zener mechanism~\cite{wittig_landauzener_2005}, in which two states, $\spindown$ and $\spinup$, are coupled with strength $\Omega/2$ while their energy separation $\Delta$ is swept through zero. The Hamiltonian for such a two-level system is

\begin{equation}
    \label{eqn:LZ_hamiltonian}
    H = \frac{1}{2}
    \begin{pmatrix}
        -\Delta(t) & \Omega \\
        \Omega & \Delta(t)
    \end{pmatrix}
\end{equation}
with $\Delta(t) = kt$ and constant coupling $\Omega$. A similar form describes a driven two-level system under the rotating wave approximation.

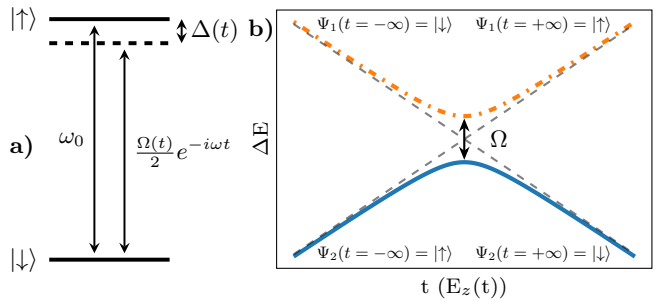
\begin{figure}
    \vspace{-0.5em}
    \begin{minipage}[c]{0.22\linewidth}
        \begin{tikzpicture}[scale=0.8]

\def\len{2}
\def\levelwidth{1.5}
\def\linewidth{0.75}
\def\ygnd{0}
\def\yexc{4}
\def\detuning{0.4}
\def\arrowshift{0.25}
\def\yarrowoffset{3}
\def\labelxoffset{0.7}

\draw[line width=\levelwidth pt] (0,\ygnd) -- node (ground) {} (\len, \ygnd);
\draw[line width=\levelwidth pt] (0,\yexc) -- node (excited) {} (\len, \yexc);
\draw[dashed, line width=\levelwidth pt] (0,\yexc-\detuning) -- node (detuned) {} (\len, \yexc-\detuning);

\draw[<->, >=stealth, line width=\linewidth pt]
  (\len+0.2, \yexc) -- (\len+0.2, \yexc-\detuning) node[midway,right] {$\Delta(t)$};

\coordinate (start) at ($(ground) + (+\arrowshift, +\yarrowoffset pt)$);
\coordinate (end) at ($(detuned) + (+\arrowshift, -\yarrowoffset pt)$);
\draw[<->, >=stealth, line width=\linewidth pt]
  (start) -- (end) node[midway,right] {$\frac{\Omega(t)}{2}e^{-i\omega t}$};

\coordinate (start) at ($(ground) + (-\arrowshift, +\yarrowoffset pt)$);
\coordinate (end) at ($(excited) + (-\arrowshift, -\yarrowoffset pt)$);
\draw[<->, >=stealth, line width=\linewidth pt]
  (start) -- (end) node[midway,left] {$\omega_0$};

\node[left=\labelxoffset of excited] {$\spinup$};
\node[left=\labelxoffset of ground] {$\spindown$};

\node[anchor=north west, yshift=5pt] at ($(current bounding box.north west)!0.5!(current bounding box.south west)$) {\textbf{a)}};
    
\end{tikzpicture}
    \end{minipage}
    \hfill
    \begin{minipage}[c]{0.76\linewidth}
        \raggedleft
        \vspace{1.5em}
\begin{tikzpicture}

\definecolor{darkorange25512714}{RGB}{255,127,14}
\definecolor{steelblue31119180}{RGB}{31,119,180}

\begin{axis}[
label style={font=\footnotesize},
tick label style={font=\footnotesize},
tick pos=left,
ticks=none,
xlabel={t (E\(\displaystyle _z\)(t))},
xmin=-5.5, xmax=5.5,
ylabel={\(\displaystyle \Delta\)E},
ymin=-2.80446073247603, ymax=2.80446073247603,
width=\linewidth,
height=4.2/5.5*\linewidth,
enlarge x limits=false,    
clip=true,                 
trim axis left,
trim axis right,
]
\addplot [ultra thick, steelblue31119180]
table {%
-5 -2.54950975679639
-4.9 -2.50049995001
-4.8 -2.45153013442625
-4.7 -2.40260275534679
-4.6 -2.35372045918796
-4.5 -2.30488611432322
-4.4 -2.2561028345357
-4.3 -2.20737400546441
-4.2 -2.15870331449229
-4.1 -2.11009478460092
-4 -2.06155281280883
-3.9 -2.01308221391974
-3.8 -1.96468827043885
-3.7 -1.91637678967368
-3.6 -1.86815416922694
-3.5 -1.82002747232013
-3.4 -1.77200451466693
-3.3 -1.72409396495667
-3.2 -1.67630546142402
-3.1 -1.62864974749023
-3 -1.58113883008419
-2.9 -1.5337861650178
-2.8 -1.48660687473185
-2.7 -1.4396180048888
-2.6 -1.39283882771841
-2.5 -1.34629120178363
-2.4 -1.3
-2.3 -1.25399362039844
-2.2 -1.20830459735946
-2.1 -1.1629703349613
-2 -1.11803398874989
-1.9 -1.07354552767919
-1.8 -1.0295630140987
-1.7 -0.986154146165801
-1.6 -0.94339811320566
-1.5 -0.901387818865997
-1.4 -0.860232526704263
-1.3 -0.820060973342836
-1.2 -0.781024967590665
-1.1 -0.743303437365925
-1 -0.707106781186548
-0.899999999999999 -0.672681202353685
-0.8 -0.640312423743285
-0.7 -0.610327780786685
-0.6 -0.58309518948453
-0.5 -0.559016994374947
-0.399999999999999 -0.53851648071345
-0.3 -0.522015325445527
-0.199999999999999 -0.509901951359278
-0.0999999999999996 -0.502493781056044
0 -0.5
0.100000000000001 -0.502493781056044
0.2 -0.509901951359279
0.300000000000001 -0.522015325445528
0.4 -0.53851648071345
0.5 -0.559016994374947
0.600000000000001 -0.58309518948453
0.7 -0.610327780786685
0.800000000000001 -0.640312423743285
0.9 -0.672681202353686
1 -0.707106781186548
1.1 -0.743303437365925
1.2 -0.781024967590666
1.3 -0.820060973342837
1.4 -0.860232526704263
1.5 -0.901387818865997
1.6 -0.943398113205661
1.7 -0.986154146165801
1.8 -1.0295630140987
1.9 -1.07354552767919
2 -1.11803398874989
2.1 -1.1629703349613
2.2 -1.20830459735946
2.3 -1.25399362039845
2.4 -1.3
2.5 -1.34629120178363
2.6 -1.39283882771841
2.7 -1.4396180048888
2.8 -1.48660687473185
2.9 -1.5337861650178
3 -1.58113883008419
3.1 -1.62864974749023
3.2 -1.67630546142402
3.3 -1.72409396495667
3.4 -1.77200451466694
3.5 -1.82002747232013
3.6 -1.86815416922694
3.7 -1.91637678967368
3.8 -1.96468827043885
3.9 -2.01308221391974
4 -2.06155281280883
4.1 -2.11009478460092
4.2 -2.15870331449229
4.3 -2.20737400546441
4.4 -2.2561028345357
4.5 -2.30488611432322
4.6 -2.35372045918796
4.7 -2.40260275534679
4.8 -2.45153013442625
4.9 -2.50049995001
5 -2.54950975679639
};
\addplot [ultra thick, darkorange25512714, dash pattern=on 1pt off 3pt on 3pt off 3pt]
table {%
-5 2.54950975679639
-4.9 2.50049995001
-4.8 2.45153013442625
-4.7 2.40260275534679
-4.6 2.35372045918796
-4.5 2.30488611432322
-4.4 2.2561028345357
-4.3 2.20737400546441
-4.2 2.15870331449229
-4.1 2.11009478460092
-4 2.06155281280883
-3.9 2.01308221391974
-3.8 1.96468827043885
-3.7 1.91637678967368
-3.6 1.86815416922694
-3.5 1.82002747232013
-3.4 1.77200451466693
-3.3 1.72409396495667
-3.2 1.67630546142402
-3.1 1.62864974749023
-3 1.58113883008419
-2.9 1.5337861650178
-2.8 1.48660687473185
-2.7 1.4396180048888
-2.6 1.39283882771841
-2.5 1.34629120178363
-2.4 1.3
-2.3 1.25399362039844
-2.2 1.20830459735946
-2.1 1.1629703349613
-2 1.11803398874989
-1.9 1.07354552767919
-1.8 1.0295630140987
-1.7 0.986154146165801
-1.6 0.94339811320566
-1.5 0.901387818865997
-1.4 0.860232526704263
-1.3 0.820060973342836
-1.2 0.781024967590665
-1.1 0.743303437365925
-1 0.707106781186548
-0.899999999999999 0.672681202353685
-0.8 0.640312423743285
-0.7 0.610327780786685
-0.6 0.58309518948453
-0.5 0.559016994374947
-0.399999999999999 0.53851648071345
-0.3 0.522015325445527
-0.199999999999999 0.509901951359278
-0.0999999999999996 0.502493781056044
0 0.5
0.100000000000001 0.502493781056044
0.2 0.509901951359279
0.300000000000001 0.522015325445528
0.4 0.53851648071345
0.5 0.559016994374947
0.600000000000001 0.58309518948453
0.7 0.610327780786685
0.800000000000001 0.640312423743285
0.9 0.672681202353686
1 0.707106781186548
1.1 0.743303437365925
1.2 0.781024967590666
1.3 0.820060973342837
1.4 0.860232526704263
1.5 0.901387818865997
1.6 0.943398113205661
1.7 0.986154146165801
1.8 1.0295630140987
1.9 1.07354552767919
2 1.11803398874989
2.1 1.1629703349613
2.2 1.20830459735946
2.3 1.25399362039845
2.4 1.3
2.5 1.34629120178363
2.6 1.39283882771841
2.7 1.4396180048888
2.8 1.48660687473185
2.9 1.5337861650178
3 1.58113883008419
3.1 1.62864974749023
3.2 1.67630546142402
3.3 1.72409396495667
3.4 1.77200451466694
3.5 1.82002747232013
3.6 1.86815416922694
3.7 1.91637678967368
3.8 1.96468827043885
3.9 2.01308221391974
4 2.06155281280883
4.1 2.11009478460092
4.2 2.15870331449229
4.3 2.20737400546441
4.4 2.2561028345357
4.5 2.30488611432322
4.6 2.35372045918796
4.7 2.40260275534679
4.8 2.45153013442625
4.9 2.50049995001
5 2.54950975679639
};
\addplot [thick, black, opacity=0.5, dashed]
table {%
-5 -2.5
-4.9 -2.45
-4.8 -2.4
-4.7 -2.35
-4.6 -2.3
-4.5 -2.25
-4.4 -2.2
-4.3 -2.15
-4.2 -2.1
-4.1 -2.05
-4 -2
-3.9 -1.95
-3.8 -1.9
-3.7 -1.85
-3.6 -1.8
-3.5 -1.75
-3.4 -1.7
-3.3 -1.65
-3.2 -1.6
-3.1 -1.55
-3 -1.5
-2.9 -1.45
-2.8 -1.4
-2.7 -1.35
-2.6 -1.3
-2.5 -1.25
-2.4 -1.2
-2.3 -1.15
-2.2 -1.1
-2.1 -1.05
-2 -1
-1.9 -0.95
-1.8 -0.9
-1.7 -0.85
-1.6 -0.8
-1.5 -0.75
-1.4 -0.7
-1.3 -0.65
-1.2 -0.6
-1.1 -0.55
-1 -0.5
-0.899999999999999 -0.45
-0.8 -0.4
-0.7 -0.35
-0.6 -0.3
-0.5 -0.25
-0.399999999999999 -0.2
-0.3 -0.15
-0.199999999999999 -0.0999999999999996
-0.0999999999999996 -0.0499999999999998
0 0
0.100000000000001 0.0500000000000003
0.2 0.1
0.300000000000001 0.15
0.4 0.2
0.5 0.25
0.600000000000001 0.3
0.7 0.35
0.800000000000001 0.4
0.9 0.45
1 0.5
1.1 0.55
1.2 0.6
1.3 0.65
1.4 0.7
1.5 0.75
1.6 0.8
1.7 0.85
1.8 0.9
1.9 0.95
2 1
2.1 1.05
2.2 1.1
2.3 1.15
2.4 1.2
2.5 1.25
2.6 1.3
2.7 1.35
2.8 1.4
2.9 1.45
3 1.5
3.1 1.55
3.2 1.6
3.3 1.65
3.4 1.7
3.5 1.75
3.6 1.8
3.7 1.85
3.8 1.9
3.9 1.95
4 2
4.1 2.05
4.2 2.1
4.3 2.15
4.4 2.2
4.5 2.25
4.6 2.3
4.7 2.35
4.8 2.4
4.9 2.45
5 2.5
};
\addplot [thick, black, opacity=0.5, dashed]
table {%
-5 2.5
-4.9 2.45
-4.8 2.4
-4.7 2.35
-4.6 2.3
-4.5 2.25
-4.4 2.2
-4.3 2.15
-4.2 2.1
-4.1 2.05
-4 2
-3.9 1.95
-3.8 1.9
-3.7 1.85
-3.6 1.8
-3.5 1.75
-3.4 1.7
-3.3 1.65
-3.2 1.6
-3.1 1.55
-3 1.5
-2.9 1.45
-2.8 1.4
-2.7 1.35
-2.6 1.3
-2.5 1.25
-2.4 1.2
-2.3 1.15
-2.2 1.1
-2.1 1.05
-2 1
-1.9 0.95
-1.8 0.9
-1.7 0.85
-1.6 0.8
-1.5 0.75
-1.4 0.7
-1.3 0.65
-1.2 0.6
-1.1 0.55
-1 0.5
-0.899999999999999 0.45
-0.8 0.4
-0.7 0.35
-0.6 0.3
-0.5 0.25
-0.399999999999999 0.2
-0.3 0.15
-0.199999999999999 0.0999999999999996
-0.0999999999999996 0.0499999999999998
0 -0
0.100000000000001 -0.0500000000000003
0.2 -0.1
0.300000000000001 -0.15
0.4 -0.2
0.5 -0.25
0.600000000000001 -0.3
0.7 -0.35
0.800000000000001 -0.4
0.9 -0.45
1 -0.5
1.1 -0.55
1.2 -0.6
1.3 -0.65
1.4 -0.7
1.5 -0.75
1.6 -0.8
1.7 -0.85
1.8 -0.9
1.9 -0.95
2 -1
2.1 -1.05
2.2 -1.1
2.3 -1.15
2.4 -1.2
2.5 -1.25
2.6 -1.3
2.7 -1.35
2.8 -1.4
2.9 -1.45
3 -1.5
3.1 -1.55
3.2 -1.6
3.3 -1.65
3.4 -1.7
3.5 -1.75
3.6 -1.8
3.7 -1.85
3.8 -1.9
3.9 -1.95
4 -2
4.1 -2.05
4.2 -2.1
4.3 -2.15
4.4 -2.2
4.5 -2.25
4.6 -2.3
4.7 -2.35
4.8 -2.4
4.9 -2.45
5 -2.5
};
\draw[<->, >={Stealth[length=5pt, width=4pt]}, thick, draw=black] (axis cs:0,-0.45) -- (axis cs:0,0.45);
\pgfmathsetmacro{\xoffset}{4.5}
\pgfmathsetmacro{\yoffset}{2.8}
\draw (axis cs:0.5,0) node[
  scale=1.0,
  anchor=west,
  text=black,
  rotate=0.0
]{$\Omega$};
\draw (axis cs:-\xoffset,\yoffset) node[
  scale=0.7,
  anchor=north west,
  text=black,
  rotate=0.0
]{$\Psi_1(t = -\infty) =  \left|\downarrow\right\rangle$};
\draw (axis cs:\xoffset,\yoffset) node[
  scale=0.7,
  anchor=north east,
  text=black,
  rotate=0.0
]{$\Psi_1(t = +\infty) =  \left|\uparrow\right\rangle$};
\draw (axis cs:-\xoffset,-\yoffset) node[
  scale=0.7,
  anchor=south west,
  text=black,
  rotate=0.0
]{$\Psi_2(t = -\infty) =  \left|\uparrow\right\rangle$};
\draw (axis cs:\xoffset,-\yoffset) node[
  scale=0.7,
  anchor=south east,
  text=black,
  rotate=0.0
]{$\Psi_2(t = +\infty) =  \left|\downarrow\right\rangle$};

\end{axis}

\node[anchor=north west, xshift=-2pt] at (current bounding box.north west) {\textbf{b)}};

\end{tikzpicture}
        \hspace{-0.7em}
    \end{minipage}
    \caption{\textbf{a)} A two-level system with energy splitting $\omega_0$, where the states are coupled by an oscillating field with frequency $\omega(t)$ and a Rabi rate $\Omega(t)/2$. In the rotating frame, the Hamiltonian simplifies to the form shown in Eqn.~\eqref{eqn:LZ_hamiltonian}, incorporating time-dependent Rabi rate $\Omega(t)$ and non-linear detuning $\Delta(t)$. Under the rotating wave approximation, this Hamiltonian also describes microwave-driven rotational transitions in TlF. \textbf{b)} Eigenenergies of the Landau-Zener system as the detuning $\Delta(t)$, here due to a time-varying E$_z$(t), evolves from $\Delta \to -\infty$ ($t \to -\infty$) to $\Delta \to +\infty$ ($t \to +\infty$). The ground state transitions from $\Psi(t \to -\infty) = \ket{\uparrow}$ to $\Psi(t \to +\infty) = \ket{\downarrow}$. Under adiabatic conditions ($\frac{d\Delta}{dt} \ll \Omega^2$), the system follows the instantaneous eigenstates, remaining in the ground state if initially prepared there.}
    \label{fig:LZ_diagram}
\end{figure}

The time evolution of the system (Fig.~\ref{fig:LZ_diagram}) depends on the rate at which the detuning $\Delta(t)$ varies relative to the coupling strength $\Omega$. For the Landau--Zener model, the final state can be solved analytically~\cite{wittig_landauzener_2005}, making it a useful benchmark for more general situations. Different regimes are characterized by the rapidity parameter  
\begin{equation}
    \Gamma = \frac{\tfrac{d\Delta}{dt}}{\Omega^2}.
\end{equation}  
In the adiabatic limit ($\Gamma \ll 1$), the state vector follows the instantaneous eigenstates, leading to complete population inversion between $\spindown$ and $\spinup$. In the sudden limit ($\Gamma \gg 1$), the system remains in its initial state $\spindown$. For intermediate values ($\Gamma \sim 1$), the final state is a coherent superposition of $\spindown$ and $\spinup$.  

Adiabatic passage corresponds to the adiabatic regime, where $\Gamma = \tfrac{d\Delta}{dt}/\Omega^2 \ll 1$. Here, $\Omega$ is typically set by the strength of an oscillating drive field (e.g., a microwave), and $\Delta$ is the detuning between the drive frequency and the transition (e.g. due to an electric field). After transforming to the rotating frame and applying the rotating wave approximation, the dynamics reduce to the Hamiltonian in Eqn.~\eqref{eqn:LZ_hamiltonian}.  

A major advantage of adiabatic passage over the conventional $\pi$-pulse method is its robustness~\cite{budker2004atomic}. Once the adiabatic condition ($\Gamma \ll 1$) is satisfied, population inversion occurs with nearly unit efficiency. This insensitivity to variations in drive strength and interaction time is particularly valuable in a molecular beam, where molecules experience different microwave intensities and transit times. Provided the coupling is sufficiently strong, AP ensures nearly uniform transfer efficiency across the ensemble.

A beam of cold TlF molecules is produced using a cryogenic buffer-gas source~\cite{Grasdijk_2021,hutzler2012buffer}, which provides high flux, low forward velocity, narrow velocity spread, and low rotational temperature. For \CENTREX, neon cooled to $19\kelvin$ serves as the buffer gas, offering higher repetition rates and more stable flux for heavy species such as TlF compared to helium~\cite{C1CP20901A}. After further cooling in the free expansion of the neon flow, the TlF beam reaches a rotational temperature of \rottemp~\cite{Grasdijk_2021}. At this temperature only $\sim5\%$ of the molecules occupy the lowest rotational state $J=0$, and just one quarter of that fraction is in the absolute ground state $\ket{J=0,F=0}$. To increase this population, rotational cooling is applied: a combination of microwave and laser fields transfers population from all hyperfine sublevels of the three lowest excited rotational levels ($J=1,2,3$) into the $J=0$ hyperfine manifold, enhancing the $\ket{J=0,F=0}$ population by a factor of $\approx 20$~\cite{vb31-f2c3}.

To collimate the diverging molecular beam, an electrostatic quadrupole lens (EQL) will be employed. This requires molecules to occupy a state with a positive quadratic Stark shift, which is realized in $\ket{J=2,m_J=0}$. Population transfer from $\ket{J=0,F=0}$ to particular hyperfine sublevels of $\ket{J=2,m_J=0}$  can be implemented via two sequential microwave-driven adiabatic passages. The necessary time-varying detuning for AP is provided by the second-order DC Stark effect: as molecules traverse a spatially varying DC electric field, the resonance frequencies of the rotational transitions shift accordingly.  

In \CENTREX, the goal is to transfer population between selected hyperfine levels of different rotational states. Microwave electric fields provide the coupling, while spatially varying DC electric fields produce the required detuning. By engineering both the magnitude and gradient of the DC field, and simultaneously applying near-resonant microwaves, robust adiabatic passage can be achieved.  

The simple Landau--Zener model is insufficient here because of the non-linear time dependence of the detuning, the spatially varying coupling strength (i.e., microwave intensity), and the presence of multiple, near-resonantly coupled hyperfine/rotational levels. Hence, we perform numerical simulations of the full TlF Hamiltonian, including levels up to $J=4$. These simulations confirmed that efficient transfer between selected hyperfine/rotational states can be achieved when the adiabatic condition $\Gamma \ll 1$ is satisfied~\cite{shore_2009}.

It is important to minimize stray microwave fields because any residual field outside the intended AP region can coherently transfer population back toward the initial state and reduce the net transfer. Additionally non-linear polarization or small transverse components in the microwave field couple hyperfine and magnetic sublevels that are not included in the designed transfer chain.

\subsection{TlF in varying electric fields}

The experiment described here performs AP in a DC electric field of $\sim100\Vcm$, with state detection at zero field. The eigenstates of TlF differ significantly between these regimes due to the combined rotational and hyperfine structure of the $\gnd,\,v=0$ ground state~\cite{Grasdijk_2021}. Rotational structure arises from the diatomic molecule’s center-of-mass rotation, while both $^{205}$Tl and $^{19}$F nuclei have spin $\sfrac{1}{2}$, producing hyperfine splittings from spin–rotation and spin–spin couplings. The effective Hamiltonian, including rotational, spin–rotation, spin–spin, and Stark interactions, is given by
\begin{equation}
    \mathcal{H}_\text{TlF} = \mathcal{H}_\text{rot} + \mathcal{H}_\text{sr} + \mathcal{H}_\text{ss} + \mathcal{H}_\text{S},
\end{equation}
with
\begin{equation} 
    \label{eq:hyperfine_hamiltonian}
    \begin{split}
        \mathcal H_\text{rot} &= B \vec{J}^2, \\
        \mathcal H_\text{sr} &= c_1(\vec I_1\cdot\vec J)+c_2(\vec I_2\cdot\vec J), \\
        \mathcal H_\text{ss} &= c_3 T^2(\vec C)\cdot T^2(\vec I_1, \vec I_2)+c_4(\vec I_1\cdot\vec I_2), \\
        \mathcal H_\text{S} &= -\vec\mu_e\cdot\vec \Evec ,
    \end{split}
\end{equation}
where constants are listed in Table~\ref{tab:hyperfine_hamiltonian}. In the ground state, TlF lacks opposite-parity near-degeneracies, so the Stark interaction appears only at second order.  

\begin{table}
    \small
    \setlength\extrarowheight{3pt}
	\centering
	\caption{Constants for rotational, hyperfine, and Stark interactions in the effective TlF ground-state Hamiltonian [Eq.~\eqref{eq:hyperfine_hamiltonian}].}
	\label{tab:hyperfine_hamiltonian}
	\begin{tabular}{r@{\hspace{5pt}}c@{\hspace{5pt}}l@{\hspace{5pt}}l | r@{\hspace{5pt}}c@{\hspace{5pt}}l@{\hspace{5pt}}l}
    	\toprule
		$B$ & = & $6.66733$ & GHz & $\mu_e$ & = & $2.1285(4)$ & MHz/V/cm \\ 
    	$c_1$ & = & $126.03(12)$ & kHz & $c_2$ & = & $17.89(15)$ & kHz \\
		$c_3$ & = & $0.70(3)$ & kHz & $c_4$ & = & $-13.30(72)$ & kHz \\
		\bottomrule
	\end{tabular}
\end{table}

At large electric fields, the Stark effect dominates. Here $J$ and $m_J$ remain good quantum numbers, and for $m_J \neq 0$ the good nuclear spin quantum numbers are $I_{Tl}=\sfrac{1}{2}, m_{Tl}$ and $I_{F}=\sfrac{1}{2}, m_{F}$. For $m_J = 0$, however, spin–spin coupling mixes the nuclear spins into singlet ($S=0$) and triplet ($S=1$) states with projection $m_S$. At low electric fields, the hierarchy reverses: $\mathcal{H}_\text{rot} \gg \mathcal{H}_\text{sr},\,\mathcal{H}_\text{ss} \gg \mathcal{H}_\text{S}$. In this regime, the good quantum numbers are $J$, $F_1 = J+I_{Tl}$, $F = F_1+I_F$, and $m_F$. 

When molecules traverse from high field to zero field, the $\ket{J,m_J=0,S=1,m_S}$ nuclear-spin triplet states adiabatically evolve into the states $\ket{J,F_1=J+1/2,F=J+1,m_F=m_S}$. The corresponding singlet state $\ket{J,m_J=0,S=0}$ evolves into $\ket{J,F_1=J+1/2,F=J,m_F=0}$. These mappings are essential for interpreting fluorescence detection, which is performed in zero field following state transfer via AP in a region of high field.

\section{Experimental Overview}
\label{sec:experimental_overview}

\begin{figure*}
	\includegraphics[width=\linewidth]{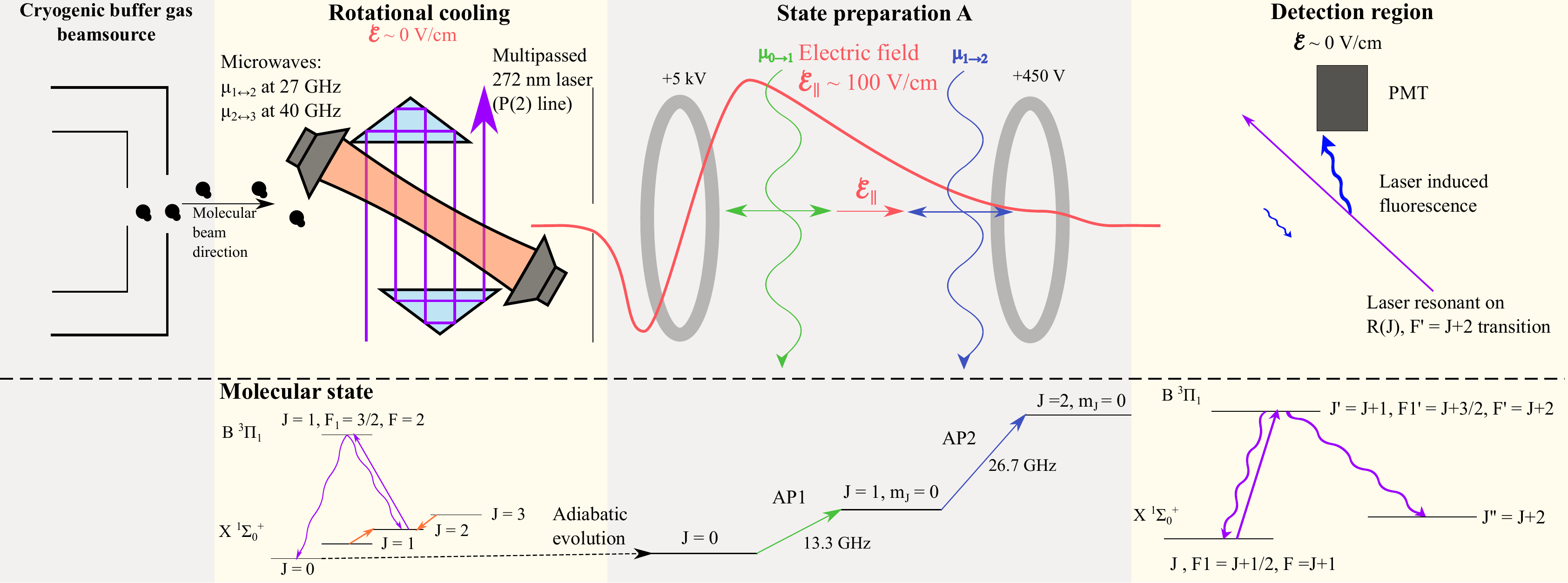}
	\caption{Schematic of the experimental apparatus. Molecules emerge from the cryogenic buffer-gas beam source and first pass through the rotational cooling region, where optical pumping and microwave transitions transfer population from the $J=1,2,3$ rotational levels to $J=0$. The molecules then enter the state preparation region, where they can be transferred to $J=1$ or $J=2$ depending on the microwave settings. Finally, the molecules are detected in the detection region by collecting laser-induced fluorescence (LIF) on a PMT. The detection laser can excite different rotational/hyperfine states, allowing determination of their relative populations under different preparation conditions. For convenience in these measurements we operate in the triplet state ($S=1$), pumping into $\ket{J=0,F=1}$ in RC. \CENTREX\ will operate in the singlet state ($S=0$), pumping into $\ket{J=0,F=0}$ in RC.}
	\label{fig:experiment_schematic}
\end{figure*}

\begin{figure}
    \centering
	\includegraphics[width=\linewidth]{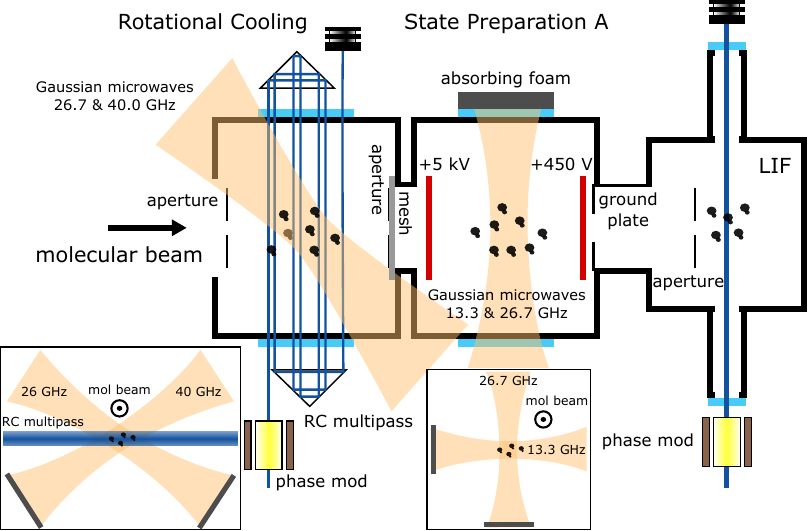}
	\caption{Schematic overview of the measurement setup. A $271.75\nm$ laser, phase-modulated on the \PtwoFtwo\ transition, makes 13 passes through the rotational cooling (RC) region, where it intersects two focused Gaussian microwave beams (inset shows frontal view). An aluminum honeycomb mesh suppresses microwave leakage from RC into SPA. In the SPA chamber, two ring electrodes at $+5\kV$ and $+450\V$ generate a spatially varying electric field, nominally parallel to the molecular beam. The molecular beam then encounters two orthogonal microwave fields that drive successive adiabatic passages. A grounded plate rapidly drives the electric field to zero at the exit of the SPA region. In the detection chamber, an aperture of $13\mm\,w\times 3\mm\,h$ restricts the transverse velocity and beam spread. Populations are measured via laser-induced fluorescence (LIF) from a single $271.75\nm$ phase-modulated probe beam, collected by two opposing photomultiplier tubes (PMTs) located above and below the plane of the page.}
	\label{fig:schematic_overview}
\end{figure}

A schematic of the apparatus is shown in Fig.~\ref{fig:experiment_schematic}. The cryogenic TlF beam, with mean forward velocity $\bar{v}_f = 184\vms$, enters the RC chamber, located $\sim40\cm$ from the source and $\sim60\cm$ from the cell exit. In this chamber, the molecules interact simultaneously with a multi-pass laser beam and two focused, free-space, nominally Gaussian-shaped microwave beams that together act to accomplish the RC (see~\cite{vb31-f2c3} for details).

\begin{figure*}
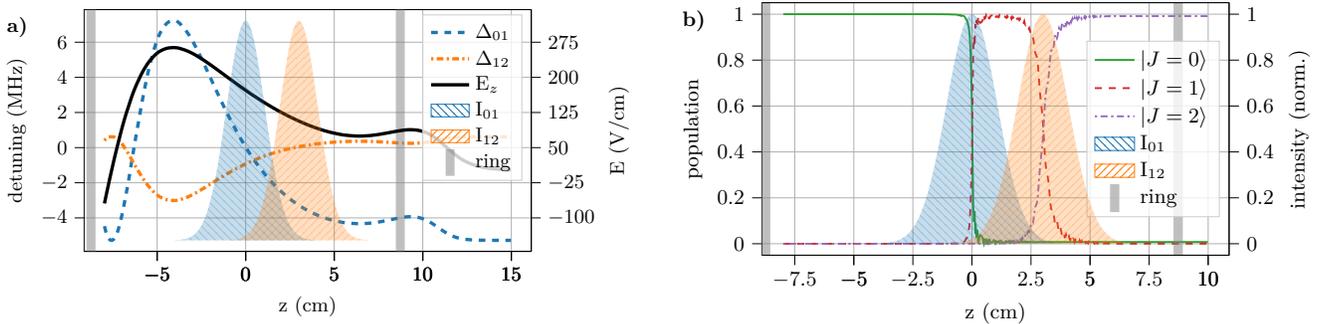

    \centering
    \small
    \begin{minipage}[c]{\columnwidth}
        \resizebox{1.02\columnwidth}{!}{
            \input{experimental_overview/figures/field_detuning}
        }
    \end{minipage}
    \hfill
    \begin{minipage}[c]{\columnwidth}
        \resizebox{1.02\columnwidth}{!}{
            \input{experimental_overview/figures/spa_simulation}
        }
    \end{minipage}
    \mbox{}
    \vspace*{-5pt}
    \caption{\textbf{a}) Applied electric field and Stark-shifted detunings of the $\ket{J=0,m_J=0}\leftrightarrow\ket{J=1,m_J=0}$ (blue) and $\ket{J=1,m_J=0}\leftrightarrow\ket{J=2,m_J=0}$ (orange) transitions ($\Delta_{01}$ and $\Delta_{02}$, respectively), as functions of the axial position $z$ in the state preparation region. The two microwave frequencies are chosen to be on resonance at the positions corresponding to the maxima of their intensities ($I_{01}$ and $I_{12}$}, respectively). The nominally Gaussian spatial profiles of the intensity of the two microwave beams are indicated by the shaded curves. The dashed black line shows the $z$-component of the electric field, $E_z$. The vertical grey lines indicate the positions of the ring electrodes in the SPA region. 
\textbf{b}) Simulated populations for molecules initially in $\ket{J=0,m_J=0}$, versus $z$. Green: $\ket{J=0,m_J=0}$, red: $\ket{J=1,m_J=0}$, purple: $\ket{J=2,m_J=0}$. Two adiabatic passages result in population inversions at the centers of the microwave intensity profiles, where the transitions cross resonance.
    \label{fig:SPA_state_evolution}
\end{figure*}

The molecules then enter the ``state preparation A'' (SPA) chamber, where the microwave AP takes place. Here, they traverse a region of spatially-varying electric field ($\mathbf{E}$) generated by two energized ring electrodes and surrounding grounded surfaces.
This DC field defines the local $z$ axis, and is nominally parallel to the direction of the molecular beam. The spatial profile of the $E$-field is designed to set the spatial evolution of Stark shifts---and hence the detunings from microwave transitions needed to accomplish AP---for the specific levels of interest. In this DC field,  at a location $30.5\cm$ downstream of the RC region, 
the molecules encounter a linearly polarized Gaussian beam of microwaves (with polarization parallel to the nominal $E$-field direction) tuned to resonance with the Stark-shifted $\ket{J=0,m_J=0}\leftrightarrow\ket{J=1,m_J=0}$ (\APone) transition, 
at a local field strength of $173\Vcm$. This is followed 
$28.6\mm$ further downstream 
by a second microwave beam, polarized in the same direction, resonant with the Stark-shifted $\ket{J=1,m_J=0}\leftrightarrow\ket{J=2,m_J=0}$ (\APtwo) transition at $110\Vcm$. Together, these microwave fields drive stepwise population transfer via AP from $\ket{J=0,m_J=0}$ to $\ket{J=1,m_J=0}$ to $\ket{J=2,m_J=0}$ (Fig.~\ref{fig:SPA_state_evolution}). The free-space microwave beams are focused at the position of the molecular beam, which they cross at right angles, orthogonally to each other and to the molecular beam.

In the Earth’s magnetic field, the nuclear-spin triplet states of $\ket{J=1,m_J=0}$ can mix with the singlet state, leading to unwanted population transfer if the electric field decays too slowly. To mitigate this, an additional grounded plate with a $44.5\mm$ diameter aperture was installed behind the $+450\V$ electrode, ensuring a rapid field drop after the SPA region and thereby suppressing such mixing.

After SPA, the molecules propagate $36.2\cm$ downstream to a region where molecular state populations are detected via laser-induced fluorescence (LIF). A $13\mm\,w \times 3\mm\,h$ aperture, positioned $25.4\mm$ upstream of the probe laser, restricts the transverse velocity distribution and spatial spread of the molecular beam. A probe laser orthogonal to the molecular beam then excites selected hyperfine levels, and the resulting fluorescence on the $B \to X$ transition, with a linewidth $\gamma_B=1.6\MHz$, is collected by two opposing photomultiplier tubes (PMTs).

\subsection{Electric field generation}
The spatially varying electric field is produced by two identical ring electrodes, with $4.5"$ major and $0.5"$ minor diameters, plus surrounding grounded sufaces. These electrodes are separated by $6.75"$ and positioned symmetrically about the SPA chamber center. The first is held at a potential of $5000\V$, and the second at $450\V$. The resulting field is shaped both to accomplish the desired adiabatic passage transfers, and to suppress unwanted Landau–Zener transitions to other, undesired molecular states. The field strength decreases from $260\Vcm$ to $100\Vcm$ over $10\cm$, corresponding to detuning rates of approximately $20\GHz/\mathrm{s}$ for \APone\ and $6\GHz/\mathrm{s}$ for \APtwo. The entrance and exit of the SPA chamber are grounded, so the $\mathbf{E}$ field outside it is nominally zero.

\subsection{Microwave generation and delivery}
The $J=0\to1$ transition occurs at $13.34\GHz$, and the $J=1\to2$ transition at $26.67\GHz$. The $13.34\GHz$ signal is generated directly by a synthesizer, while the $26.67\GHz$ signal is obtained by amplifying a $13.33\GHz$ source and frequency doubling. Both fields are delivered by cylindrically symmetric, spot-focusing lens antennas fed by circular waveguides, producing free-space beams with nearly Gaussian shapes, focused at the molecular beam position. Both microwave beam have $-3\,\mathrm{dB}$ diameters of $2.54\cm$ at the molecular beam position. The beams enter and exit the chamber through fused silica windows and are terminated in absorbing foam.

Microwave powers of $0.2\mW$ and $0.05\mW$ are used for the $J=0\to1$ (Rabi rate $\Omega_{01} \approx 2\pi \times 88\kHz$) and $J=1\to2$ (Rabi rate $\Omega_{12} \approx 2\pi \times 35\kHz$) transitions, respectively. These correspond to Rabi rates of $\Omega_{01} \approx 2\pi \times 88\kHz$ and $\Omega_{12} \approx 2\pi \times 35\kHz$, respectively.  Simulations indicate that the adiabatic transfer efficiency is robust against changes in microwave intensity, maintaining high fidelity even for molecules displaced from the beam center. The microwave beam polarizations are aligned along $\hat{z}$, to selectively drive $\Delta m_J=0$ transitions. \APone\ is centered between the ring electrodes, and \APtwo\ is located $28.6\mm$ downstream.

We find it important to suppress leakage of the powerful microwaves applied in the RC region, into the SPA chamber where the microwave AP takes place. To accomplish this, an aluminum hexagonal honeycomb mesh with $3/8''$ thickness and $1/16''$ cell size is installed where molecules exit the RC region and enter the SPA region. The cutoff behavior of the hexagonal cells can be modeled using waveguides of equal cross-sectional area with square or circular geometry, since finite-element analyses show that the fundamental eigenvalue of a regular hexagon differs by  $\leq2\%$ from that of an equal-area square guide~\cite{Vaish2010HexagonalWaveguide} and by only $\sim\,0.5\%$ from an equal-area circular guide~\cite{6328900}. For the $1/16''$ hexagonal cells used here, this corresponds to an effective cutoff frequency of $f_{c,11}\approx106~\mathrm{GHz}$ for the dominant TE$_{11}$ mode, well above the RC microwave frequencies. For a $3/8''$-long cell, transmission at $26.6~\mathrm{GHz}$ is expected to be attenuated by more than $180~\mathrm{dB}$. A bench test of the honeycomb mesh, however, showed only $60~\mathrm{dB}$ attenuation. Additional suppression of stray microwave fields is provided by mounting Laird Eccosorb HR microwave-absorbing foam (which reduces reflection by $\geq20~\mathrm{dB}$) at the microwave exit windows in the RC region.

\subsection{Detection Laser}
Detection is performed on R-branch transitions, where a line labeled $R(J)$ excites a ground rotational level $\ket{J}$ to an excited level $\ket{J'=J+1}$. We use these transitions to probe population from the $J=0,1,2$ rotational levels of the $X^1\Sigma_0^+$ ground state.
All transitions lie at a wavelength of $271.75\nm$ and are readily saturated, yielding about two spontaneously emitted photons per molecule before pumping into a dark rotational state \cite{norrgard_hyperfine_2017,meijer_2020,dissertation_timgren_2023}. The ultraviolet light is generated by frequency quadrupling $1087\nm$ radiation in two successive stages of second harmonic generation, producing up to $40\mW$ at $271.75\nm$. The probe beam is expanded to elliptical $1/e^2$ diameters of $3.8\mm$ (width) and $6.2\mm$ (height), corresponding to a peak intensity of $440\mW/\cm^2$. To efficiently cover the transverse Doppler distribution of the molecular beam, the laser is phase-modulated with an electro-optic modulator at frequency $\gamma_B$ and modulation index $\beta=3.8$.

\subsection{Measurement scheme for population transfer efficiency}
In the $X^1\Sigma_0^+$ ground state, hyperfine splittings are $\sim\!100\kHz$, much smaller than the natural linewidth of optical transitions to the $B^3\Pi_1$ state, $\gamma_B=1.6\MHz$. All ground-state hyperfine levels therefore are typically simultaneously resonant at a single laser frequency. By contrast, hyperfine splittings in the $B^3\Pi_1$ state are typically $\sim100\MHz$ or larger, so the excited-state substructure is fully resolved. This situation couples many ground states with fewer excited states; when the number of ground states exceeds the number of excited states, coherent dark states can form, obscuring the relation between the fluorescence signal and molecular population.

To avoid dark-state formation and provide a clean measure of populations, we probe using only using laser frequencies resonant with the transition to the highest-$F'$ hyperfine level of each excited-state rotational manifold, namely $\ket{\tilde{J}'=J+1,\,F'=J'+1 =J+2}$. Selection rules ensure that this laser frequency excites only population in the highest-$F$ hyperfine level of each ground-state rotational manifold, i.e., the level with $F=J+1$. Here, $2F+1 = 2J+3$ ground state Zeeman sublevels couple to $2F'+1 = 2J+5$ excited state sublevels, eliminating dark states and ensuring that scattering continues until the $F=J+1$ level is completely pumped into other states.

While convenient for detecting populations after the microwave AP stages, these probe transitions address only the nuclear-spin triplet manifold ($S=1$), not the singlet state ($S=0$). However, \CENTREX\,will operate in the $S=0$ state to provide a well-defined initial state for the Schiff moment measurement. Hence it is important to relate the state transfer efficiency measurements reported here---performed entirely in the $S=1$ manifold---to the efficiencies in the $S=0$ manifold that are directly relevant to \CENTREX\.
We find that the transfer efficiencies are independent of $S$, which makes sense since the microwaves do not couple to the nuclear spins (but only to molecular rotation). Full numerical simulations were performed to confirm that the distinction between $S=0$ and $S=1$ has no measurable impact on the microwave-driven AP transfer efficiency between rotational states. 

To maximize the fluorescence signal during our tests of AP transfer efficiency (using $S=1$ states), we slightly modify the usual \CENTREX\ RC protocol \cite{vb31-f2c3}. In particular, for the measurements reported here, the optical pumping for RC is performed using the \PtwoFtwo\ transition, which serves to enhance population in the $\ket{J=0,\,F=1}$ states. In P-branch notation, a line labeled $P(J)$ connects a ground rotational level $J$ to an excited level $J'=J-1$. (By contrast, \CENTREX\ normally employs the \PtwoFone\ transition for RC, to enhance the population in the $S=0$ state $\ket{J=0,\,F=0}$.)

In the measurements described here, state transfer efficiency is defined as the fraction of molecules transferred from the initially populated states to the target states, with the SPA active.
Efficiencies are first optimized for \APone\ $(J=0\rightarrow J=1)$ and subsequently for \APtwo $(J=1\rightarrow J=2)$, keeping \APone\ fixed at its optimum.

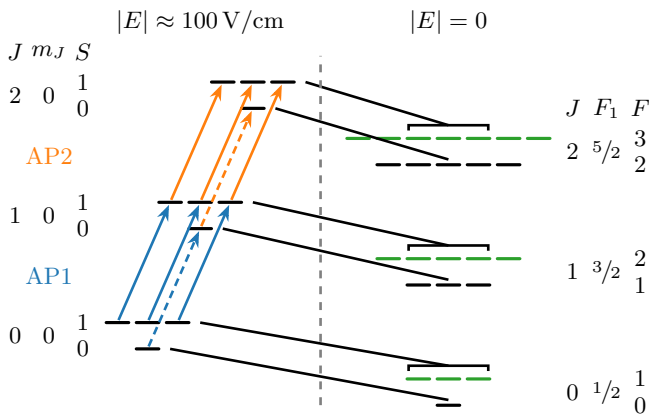
\begin{figure}
    \centering

\definecolor{darkgray176}{RGB}{176,176,176}
\definecolor{darkorange25512714}{RGB}{255,127,14}
\definecolor{lightgray204}{RGB}{204,204,204}
\definecolor{steelblue31119180}{RGB}{31,119,180}
\definecolor{forestgreen4416044}{RGB}{44,160,44}

\def\connwidth{1}      
\def\levelwidth{1.2}   
\def\levelyspacing{0.35}
\def\levelxspacing{0.4}
\def\levellen{0.3}

\def\labelxoffset{0.45}
\def\labelxshift{5.5*\levellen}
\def\labelyshift{0.4}

\def\Jxshift{0.7}
\def\Jyshift{1.6}

\def\Jzerofieldxshift{4.0}
\def\Jzerofieldyshift{-0.4}

\def\bracesep{3pt}
\def\braceheight{2pt}                               

\pgfmathsetmacro{\levelyspacingpt}{\levelyspacing*72/2.54}

\tikzset{
  level/.style   = {line width=\levelwidth pt, line cap=round},
  connect/.style = {line width=\connwidth},
  arrowA/.style  = {steelblue31119180, -{Stealth[scale=0.8]}, line width=\connwidth,
                    shorten >=-3pt, shorten <=-3pt},
  arrowB/.style  = {darkorange25512714, -{Stealth[scale=0.8]}, line width=\connwidth,
                    shorten >=-3pt, shorten <=-3pt},
  lab/.style     = {inner sep=0pt, outer sep=0pt},
}

\begin{tikzpicture}
  \def\coordinates{{0/0/0}, {1/-\levelxspacing/\levelyspacing}, {2/0/\levelyspacing}, {3/\levelxspacing/\levelyspacing}}

  \foreach \J in {0,1,2} {
    \foreach \i/\xx/\yy in \coordinates {
      \pgfmathsetmacro{\xx}{\xx + \J*\Jxshift}
      \pgfmathsetmacro{\yy}{\yy + \J*\Jyshift}
      \draw[level] (\xx,\yy) -- node (J_\J_\i) {} (\xx+\levellen,\yy);
    }
    \node[lab] at ($(J_\J_0) + (-\levellen - \Jxshift*\J - 0.8*\Jxshift - 2*\labelxoffset,\levelyspacing/2)$) {\J};
    \node[lab] at ($(J_\J_0) + (-\levellen - \Jxshift*\J - 0.8*\Jxshift - \labelxoffset,\levelyspacing/2)$) {0};
    \node[lab] at ($(J_\J_0) + (-\levellen - \Jxshift*\J - 0.8*\Jxshift,0)$) {0};
    \node[lab] at ($(J_\J_0) + (-\levellen - \Jxshift*\J - 0.8*\Jxshift,\levelyspacing)$) {1};
  }
  \node[lab] at ($(J_2_0) + (-\levellen - \Jxshift*2.8 - 2*\labelxoffset, \labelyshift + \levelyspacing)$) {$J$};
  \node[lab] at ($(J_2_0) + (-\levellen - \Jxshift*2.8 - \labelxoffset, \labelyshift + \levelyspacing)$) {$m_J$};
  \node[lab] at ($(J_2_0) + (-\levellen - \Jxshift*2.8, \labelyshift + \levelyspacing)$) {$S$};
    
    \draw[arrowA, densely dashed] (J_0_0) -- (J_1_0);
    \draw[arrowB, densely dashed] (J_1_0) -- (J_2_0);
    
    \foreach \i in {1,2,3} {
      \draw[arrowA] (J_0_\i) -- (J_1_\i);
      \draw[arrowB] (J_1_\i) -- (J_2_\i);
    }

    \coordinate (labelcol) at
      ($(J_0_0) + (-\levellen - 0.8*\Jxshift - \labelxoffset,0)$);
    
    \coordinate (labJ0) at ($(J_0_0) + (0,\levelyspacing/2)$);
    \coordinate (labJ1) at ($(J_1_0) + (0,\levelyspacing/2)$);
    \coordinate (labJ2) at ($(J_2_0) + (0,\levelyspacing/2)$);
    
    \coordinate (AP1mid) at ($(labJ0)!0.5!(labJ1)$);
    \coordinate (AP2mid) at ($(labJ1)!0.5!(labJ2)$);
    
    \node[lab] at (labelcol |- AP1mid) {\color{steelblue31119180}{AP1}};
    \node[lab] at (labelcol |- AP2mid) {\color{darkorange25512714}{AP2}};

  \foreach \i/\F in {0/1, 1/2, 2/3} {
    \pgfmathsetmacro{\minF}{int(-\F)}
    \pgfmathsetmacro{\maxF}{int(\F)}
    \foreach \mFi in {\minF,...,\maxF} {
      \pgfmathtruncatemacro{\mF}{\mFi} 
      \pgfmathsetmacro{\xx}{\Jzerofieldxshift + (\mF * \levelxspacing)}
      \pgfmathsetmacro{\yy}{\i * \Jyshift + \Jzerofieldyshift}
      \draw[level, draw=forestgreen4416044] (\xx,\yy) -- node (Jzerofield_\F_\mF) {} (\xx+\levellen,\yy);
    }
  }

  \foreach \i/\Fsub in {0/0, 1/1, 2/2} {
    \pgfmathsetmacro{\ybase}{\i*\Jyshift + \Jzerofieldyshift - \levelyspacing}
    \foreach \mFi in {-\Fsub,...,\Fsub} {
      \pgfmathtruncatemacro{\mF}{\mFi} 
      \pgfmathsetmacro{\xx}{\Jzerofieldxshift + (\mF * \levelxspacing)}
      \draw[level] (\xx,\ybase) -- node (JzerofieldSub_\Fsub_\mF) {} (\xx+\levellen,\ybase);
    }
  }

  \foreach \J/\F in {0/1, 1/2, 2/3} {
    \path let \p1 = (Jzerofield_\F_-1.west),
              \p2 = (Jzerofield_\F_1.east) in
          coordinate (brace_start) at (\x1, \y1 + \bracesep)
          coordinate (brace_end)   at (\x2, \y2 + \bracesep);
    \draw[connect] (brace_start) -- ++(0, \braceheight) -- ($(brace_end)+(0,\braceheight)$) -- ++(0,-\braceheight);
    \coordinate (brace_mid) at ($(brace_start)!0.5!(brace_end)$);
    \node[lab, above=0pt] at (brace_mid) (Jbrace_\J) {};
    \draw[connect] ($(J_\J_3)+(\levellen,0)$) -- ($(brace_mid)+(0,\braceheight)$);
   }

  \def\connectgap{2pt}
  \foreach \J/\F in {0/0, 1/1, 2/2} {
    \path let \p1 = (JzerofieldSub_\F_0.west), \p2 = (JzerofieldSub_\F_0.east) in
      coordinate (target_\J) at ({0.5*(\x1+\x2)}, {\y1+\connectgap}); 
    \draw[connect, rounded corners=0.8pt]
      ($(J_\J_0)+(\levellen,0)$) -- (target_\J);
  }

  \path let \p1 = (J_2_3), \p2 = (Jzerofield_3_-3), \p3 = (Jzerofield_1_0), \p4 = (J_2_1) in
        coordinate (midpoint_x) at ({0.5*(\x1+\x2)}, 0)
        coordinate (start_point) at ({0.5*(\x1+\x2)}, \y3 - \levelyspacingpt)
        coordinate (end_point)   at ({0.5*(\x1+\x2)}, \y4 + \levelyspacingpt);
  \draw[dashed, gray, line width=\connwidth] (start_point) -- (end_point);

  \node[lab] at ($(Jzerofield_3_0) + (\labelxshift, \labelyshift)$) {$J$};
  \node[lab] at ($(Jzerofield_3_0) + (\labelxshift+\labelxoffset, \labelyshift)$) {$F_1$};
  \node[lab] at ($(Jzerofield_3_0) + (\labelxshift+2*\labelxoffset, \labelyshift)$) {$F$};

  \foreach \i/\F in {0/1, 1/2, 2/3}{
    \pgfmathsetmacro{\val}{int(2*\F-1)}
    \pgfmathsetmacro{\valJ}{int(\F-1)}
    \pgfmathtruncatemacro{\Fsub}{\F-1}
    \coordinate (mid_\F) at ($(Jzerofield_\F_0)!0.5!(JzerofieldSub_\Fsub_0)$);
    \node[lab] at ($(mid_\F) + (\labelxshift, 0)$) {$\valJ$};
    \node[lab] at ($(mid_\F) + (\labelxshift+\labelxoffset, 0)$) {$\sfrac{\val}{2}$};
    \node[lab] at ($(Jzerofield_\F_0) + (\labelxshift+2*\labelxoffset, 0)$) {$\F$};
    \node[lab] at ($(JzerofieldSub_\Fsub_0) + (\labelxshift+2*\labelxoffset, 0)$) {$\Fsub$};
  }

  \node[lab] at ($(J_1_2)!(J_2_2)!(J_1_2) + (0, 2*\labelyshift)$) {$\left|E\right| \approx 100 \Vcm$};
  \node[lab] at ($(Jzerofield_3_0)!(J_2_2)!(Jzerofield_3_0) + (0, 2*\labelyshift)$) {$\left|E\right| = 0$};
\end{tikzpicture}
    \caption{Adiabatic evolution of $\ket{J,m_J=0,S}$ states from high field ($|E|\approx100\Vcm$) to zero field. Blue arrows indicate the $\APone$ ($J=0\to1$) transition, and orange arrows the $\APtwo$ ($J=1\to2$) transition. Nuclear singlet transitions are indicated with dotted lines and nuclear triplet transitions with solid lines. At zero field, the nuclear spin triplet states map to states (in the basis of fully coupled angular momenta) of the form $\ket{J,F_1=J+1/2,F=J+1,m_F=-1,0,1}$, while the singlet state maps to $\ket{J,F_1=J+1/2,F=J,m_F=0}$. Population in the highest $F$ manifold of each $J$ (green) state is detected via LIF.}
    \label{fig:state_evolution}
\end{figure}

Since SPA transfers population between $\ket{J,m_J=0}$ states in fields of order $100\Vcm$, it is necessary to track their adiabatic evolution as the field encountered by the molecules evolves to or from $E=0$, as is the case just before and after the SPA region. 
In the absence of magnetic fields, the triplet states map onto the states (described in the basis of fully-coupled angular momenta) of the form $\ket{J,F_1=J+1/2,F=J+1,m_F=-1,0,+1}$. These states are addressed by the detection laser. The singlet state evolves into $\ket{J,F_1=J+1/2,F=J,m_F=0}$, which is not detected here (Fig.~\ref{fig:state_evolution}).

Finally, states  outside the $\ket{J,m_J=0}$ manifold must also be considered. At zero $E$-field these are the $\ket{J,F_1=J+1/2,F=J+1,m_F\neq-1,0,+1}$ states. These states are not affected by the adiabatic passage but contribute background in detection and must be included when extracting efficiencies.

\subsubsection{Effect of population in unwanted states}
If the RC protocol worked perfectly, only states $|J=0\rangle$ would be populated as molecules enter the SPA region. However, in practice, there remains residual population in various sublevels of the $|J=1\rangle$ and $|J=2\rangle$ manifolds. The populations of certain sublevels of these manifolds are not affected by our adiabatic passage protocol, but will appear directly in the LIF signal after the SPA region. In particular, states of the form $\ket{J,F_1=J+1/2,F=J+1,m_F\neq-1,0,+1}$ in zero $E$-field have this property. 

In addition, certain sublevels in these manifolds also can be transferred ``backwards'' by the AP process itself (i.e., to states with a lower $J$ value but still with $F=J+1$ so they are detected by LIF).  This will result, for example, in nonzero detected LIF signal from the states meant to be depopulated by the AP process.  

Both these effects complicate the analysis needed to correctly translate the measured LIF signals into true state transfer efficiencies. Next we discuss the procedure we use to account for these effects. 

\subsubsection{\APone\ transfer efficiency}
We denote the efficiency of transfer in $\APone$ as $\epsone$. To determine its value, we measure LIF signals---proportional to the populations in zero $E$-field after the SPA region---from the $\ket{J=0, F=1}$ and $\ket{J=1,F=2}$ states, both with and without the $\APone$ microwaves applied. We denote these raw signals (with the obvious correspondences) as $S_{0/1}^{\rm ON/OFF}$.  The $S^{\rm OFF}_{0}$ and $S^{\rm OFF}_{1}$ signals are simply proportional to the overall photon detection efficiency $K$, the total number of Zeeman sublevels $2F+1=2J+3$, the population $n_J$ of each Zeeman sublevel at the input to the relevant AP region (assumed to all be the same within the $\ket{J,F=J+1}$ manifold), and the average number of LIF photons emitted when probing each of these manifolds, $n_{\gamma_J}$. Specifically, we write
\begin{subequations}
    \begin{align}
        S_0^{\mathrm{OFF}} &=  3 n_{\gamma_0} n_0 K, \\
        S_1^{\mathrm{OFF}} &= 5 n_{\gamma_1} n_1 K.
    \end{align}
\end{subequations}
We expect $n_1 \ll n_0$ after rotational cooling.

When the $\APone$ microwaves are applied, we detect the analogous signals, $S_{0,1}^{\mathrm{ON}}$. The expected size of these signals must account for both the desired transfer $J=0\to J=1$ and the undesired backward transfer $J=1\to J=0$ (arising from unwanted initial population in the $\ket{J=1,F=2}$ manifold), both occurring with efficiency $\epsone$. This leads to the expressions
\begin{subequations}
    \begin{align}
        S_0^{\mathrm{ON}} &= 3 n_{\gamma_0} \bigl[(1-\epsone) n_0 + \epsone n_1 \bigr] K, \label{eq:S0ON} \\
        S_1^{\mathrm{ON}} &= n_{\gamma_1} \bigl[5 n_1 + 3 \epsone (n_0 - n_1) \bigr] K, \label{eq:S1ON}
    \end{align}
\end{subequations}
In Eq.~\eqref{eq:S0ON}, the first term comes from incomplete transfer of $\ket{J=0}$ population into $\ket{J=1}$, and the second term from backward transfer from residual $\ket{J=1}$ population. In Eq.~\eqref{eq:S1ON}, the first term comes from the residual population in $\ket{J=1}$, and the second term accounts both for successful transfer from $\ket{J=0}$ and backwards transfer from residual $\ket{J=1}$ population.

To eliminate the poorly known quantities $K$ and $n_J$, it is convenient to define ratios of the signals as follows:
\begin{subequations}
    \begin{align}
        R_0 =&~ S_0^{\mathrm{ON}}/S_0^{\mathrm{OFF}}, \\
        R_{10}^{\rm OFF} =&~ S_1^{\mathrm{OFF}}/S_0^{\mathrm{OFF}}, \\
        R_{\rm{t}} =&~ S_1^{\mathrm{ON}}/S_0^{\mathrm{OFF}}.
    \end{align}
\end{subequations}
Note that for high efficiency and low unwanted populations, we expect $R_0, R_{10}^{\rm OFF} \ll 1$, and $R_{\rm{t}} \sim n_{\gamma_1}/n_{\gamma_0}$.

The values of $n_{\gamma_J}$ can be calculated from known branching ratios \cite{norrgard_hyperfine_2017,meijer_2020,dissertation_timgren_2023} for decay of the $\ket{J'=J+1,F'=J'+1}$ states excited by the laser into different hyperfine/rotational levels of the ground state (in the approximation that the probe transitions are fully saturated, which we find is well satisfied).  We find $n_{\gamma_0} = \nphotonRzeroFtwo$, $n_{\gamma_1} = \nphotonRoneFthree$, and $n_{\gamma_2} = \nphotonRtwoFfour$.

With the measured ratios of signals, there are two distinct ways to determine the efficiency $\epsone$ from our data. One uses $R_0$, which nominally corresponds to the fractional removal of $\ket{J=0,F=1}$ population by the $\APone$ microwaves; the other uses $R_t$, which nominally corresponds to the fraction of the original population in $\ket{J=0}$ that is transferred into $\ket{J=1}$.  We hence define two different ``apparent'' efficiencies, derived from these different ratios and corresponding to removal of population from $\ket{J=0}$ ($\epsone^{\rm \dep}$), or appearance of population in $\ket{J=1}$ ($\epsone^{\rm \acc}$).  These are related to the signal ratios via the relationships
\begin{subequations}
    \begin{align}
        \epsone^{\mathrm{\dep}} &= 
        \left(1 - R_0\right)\Big/\left(1 - \tfrac{3 n_{\gamma_0} }{5 n_{\gamma_1} } R_{10}^\mathrm{OFF}\right), 
        \label{eq:eps_one_dep} \\
        \epsone^{\mathrm{\acc}} &= 
        \frac{n_{\gamma_0}}{n_{\gamma_1}} \left( R_{\rm{t}} - R_{10}^\mathrm{OFF} \right) 
        \Big/ \left(1 - \tfrac{3 n_{\gamma_0} }{5 n_{\gamma_1}} R_{10}^\mathrm{OFF} \right).
        \label{eq:eps_one_acc}
    \end{align}
\end{subequations}
Both methods should give consistent values of $\epsone$ if all population removed from $\ket{J=0,\,F=1}$ is ultimately transferred into $\ket{J=1,\,F=2}$ states.  Writing these relationships in terms of the ratios makes it evident that in the limit where $R_{10}^{\rm OFF} \rightarrow 0$, $R_{0} \rightarrow 0$ and and $R_t \rightarrow n_{\gamma_1} / n_{\gamma_0}$, $\epsone^{\mathrm{\dep}} = \epsone^{\mathrm{\acc}} = 1$.

\subsubsection{\APtwo\ transfer efficiency}

Extracting the $\APtwo$ transfer efficiency requires probing the populations in $\ket{J=1,\,F=2}$ and $\ket{J=2,\,F=3}$. We probe $\ket{J=1,\,F=2}$ as for $\APone$. For $\ket{J=2,\,F=3}$, we tune the probe laser to excite the $\ket{\tilde{J}'=3,\,F'=4}$ state using the \RtwoFfour\ transition.

Including residual $J=2$ population after the RC region and finite \APone\ efficiency, the measured signals $\Sigma$ (defined in analogy with the signals for $\APone$) are given by
\begin{subequations}
    \begin{align}
        \Sigma_1^{\mathrm{OFF}} &= S_1^\mathrm{ON} = n_{\gamma_1} \left[5 n_1 + 3\epsone(n_0 - n_1)\right]K \\
         \Sigma_1^{\mathrm{ON}} &= 
             n_{\gamma_1} \left(3\epstwo n_2 + 2n_1\right) \\
            &~~+ 3n_{\gamma_1}\left[1-\epstwo\left(\epsone n_0 + n_1(1-\epsone)\right)\right] K
         \\
        \Sigma_2^{\mathrm{OFF}} &= 7 n_{\gamma_2} n_2 K \\
        \Sigma_2^{\mathrm{ON}} &= 
            n_{\gamma_2}\left[4n_2 + 3\epstwo\left(\epsone n_0 + n_1(1-\epsone)\right)\right] \\
        &~~+ 3n_{\gamma_2}n_2(1-\epstwo) K,
    \end{align}
\end{subequations}
where $\epstwo$ is the $\ket{J=1}\rightarrow\ket{J=2}$ transfer efficiency due to the $\APtwo$ microwaves. Note that here, ``ON/OFF'' corresponds to the $\APtwo$ microwaves; for all these measurements, $\APone$ is operated at maximum transfer efficiency.  

For simplicity, we combine these signal relations with only the expression for $\epsilon_{01}^{\mathrm{\acc}}$ [Eq.~\eqref{eq:eps_one_acc}] to find the following expressions for the apparent $\APtwo$ transfer efficiencies:

\begin{subequations}
    \begin{align}
\epstwo^{\mathrm{\dep}} &= \frac{35 \, n_{\gamma_2} \left(\Sigma_{1}^{\mathrm{OFF}} - \Sigma_{1}^{\mathrm{ON}}\right)}%
{d}, 
\label{eq:eps_two_dep} \\
    \epstwo^{\mathrm{\acc}} &= \frac{35 \, n_{\gamma_1} \left(\Sigma_{2}^{\mathrm{ON}} - \Sigma_{2}^{\mathrm{OFF}}\right)}%
{d},
        \label{eq:eps_two_acc}
    \end{align}
\end{subequations}
where the common divisor is
\begin{equation}
    \begin{split}
        d &= 35 \, S_{1}^{\mathrm{ON}} \, n_{\gamma_2} - 14 \, S_{1}^{\mathrm{OFF}} \, n_{\gamma_2} - 15 \, n_{\gamma_1} \, \Sigma_{2}^{\mathrm{OFF}}.
    \end{split}
\end{equation}
Note that we have not found a form of these equations that makes their limiting forms easily apparent, so we leave them in terms of the raw signals.

\subsubsection{Total SPA transfer efficiency}

The total $\ket{J=0}\rightarrow\ket{J=2}$ transfer efficiency in the SPA region, $\epstot$, is given by by the product of the \APone\ and \APtwo\ efficiencies: $\epstot=\epsone \epstwo$.
The apparent value(s) of this quantity can be related to the various signals using the expressions above. 

\section{Results}
\label{sec:results}
\begin{figure*}
    \centering
    \small
    \begin{minipage}[t]{\columnwidth}
        \input{results/figures/J01_transfer_efficiency_rc_scan}
    \end{minipage}
    \hfill
    \begin{minipage}[t]{\columnwidth}
        \input{results/figures/J12_transfer_efficiency}
    \end{minipage}
    \mbox{}
    \vspace{-10pt}
    \caption{Apparent transfer efficiencies based on \depletion and \accumulation measurements of the $\APone$ $J=0\to1$ and $\APtwo$ $J=1\to2$ microwave transitions. In both panels, error bars indicate the $95\%$ confidence level, and the connecting lines serve only as visual guides. The $26.669\GHz$ offset corresponds to the RC J12 microwave resonance. \textbf{a)} Apparent efficiencies for the $\APone$ $J=0\to1$ transfer, versus the detuning from resonance of the high-power $J=1\to2$ microwaves in the RC region. This clearly shows improved apparent efficiency in \accumulation when the RC J12 microwaves are detuned from resonance. With the RC microwaves on resonance, the apparent \depletion and \accumulation efficiencies are very different: $0.97(1)$ and $0.23(2)$, respectively. When the RC J12 microwaves are detuned by $-7.4~\MHz$, the apparent transfer efficiencies are instead consistent with each other (as expected): $\epsone^{\mathrm{\dep}} = \epsonedep$ and $\epsone^{\mathrm{\acc}} = \epsoneacc$.  The efficiency of RC was not noticeably changed with this detuning applied. We thus set the RC J12 microwave detuning at $-7.4~\MHz$ for subsequent characterization of $\APtwo$ efficiencies. \textbf{b)} Apparent efficiencies for $\APtwo$ $J=1\to2$ transfer, versus detuning of the J12 microwaves in the SPA region. with maximal \depletion and \accumulation efficiencies of $\epstwo^{\mathrm{\dep}} = \epstwodep$ and $\epstwo^{\mathrm{\acc}} = \epstwoacc$, respectively. The vertical green line marks the resonance frequency of the transfer to $\ket{F1=3/2,F=1}$, rather than the intended $\ket{F1=5/2,F=3}$, consistent with Fig.~\ref{fig:J12_transitions}.}
    \label{fig:SPA_efficiency_scan}
\end{figure*}

\begin{figure}
    \centering
    \input{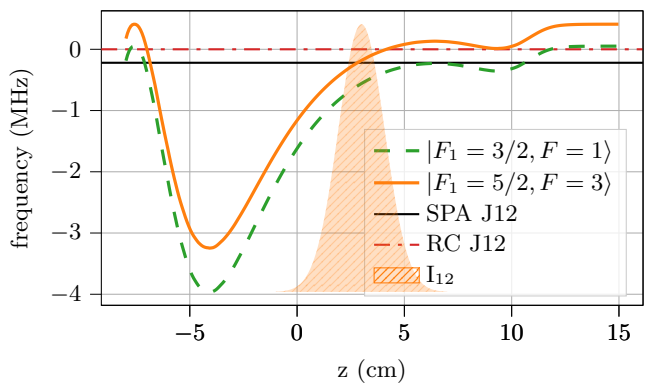}
    \vspace*{-20pt}
    \caption{Detuning of the $\ket{J=1,F_1=3/2,F=2,m_F=0}\!\to\! J=2$ transition driven by linearly polarized microwaves, offset by $26.669\GHz$; the RC J12 microwave resonance. The target state is $\ket{F_1=5/2,F=3}$. The solid horizontal line marks the $\APtwo$ frequency, and the dash-dotted line the RC $J=1\to2$ frequency. The shaded curve shows the assumed microwave intensity profile. State labels are given in the coupled basis at zero field.}
    \label{fig:J12_transitions}
\end{figure}

\subsection{\APone\ transfer efficiency}
To experimentally determine the \APone\ transfer efficiency, we initially set the RC region parameters to their usual optimized settings (i.e., both RC microwaves and the RC laser on resonance), to maximize the population detected in the $\ket{J=0,F=1}$ states. Next we determine the optimal $\APone$ microwave power, i.e., the power that minimizes the observed value of $R_0$, and the optimal $\APone$ microwave frequency, i.e., the frequency that maximizes the observed value of $R_t$.  We found, as expected, that the measured values of $R_0$ and $R_t$ were only weakly sensitive to both the power and frequency of the $\APone$ microwaves, over a fairly broad range around their expected optimum values. In practice, suitable performance is obtained for powers between $0.1\mW$ and $0.6\mW$; the current setpoint is $0.2\mW$. Similarly, varying the microwave frequency by up to $\pm 1\MHz$ around the resonance of $13.34\GHz$ does not measurably change the observed performance. The central frequency agrees with the expected value from the rotational splittings and the Stark shift.

In this configuration, we expected to find $\epsone \approx 1$---and indeed, the measured apparent \depletion efficiency was $\epsone^{\mathrm{\dep}} = 0.97(1)$.
However, the measured apparent \accumulation efficiency was much lower: $\epsone^{\mathrm{\acc}} = 0.23(2)$. 

We traced the origin of this discrepancy to leakage of the very high power microwaves from the RC region (with typical peak power of $0.5\W$) into the SPA chamber, where the typical peak \APone\ microwave power is only $0.05\mW$. Specifically, the discrepancy between apparent \depletion and \accumulation efficiencies was found to arise from population transfer from $\ket{J=1}$ to $\ket{J=2}$ in the SPA region, induced by microwaves resonant with the $J=1\rightarrow J=2$ transition at zero $E$-field that leak from the RC region into the SPA region. Note that molecules are also near resonance with this frequency at various positions in the SPA region where the $E$-field is small (see Fig.~\ref{fig:J12_transitions}).  These stray microwaves reduce the measured population of $\ket{J=1}$ after the SPA region, but do not affect the population of $\ket{J=0}$ states.  Hence, the apparent efficiency determined from \depletion\ of the $\ket{J=0}$ state can still provide sensible results, while the apparent efficiency determined from \accumulation\ into the $\ket{J=1}$ state will be smaller than the actual efficiency. Because the microwave power in the RC region is about $10^{4}$ times larger than that in the SPA region, even a very small fractional leakage  produces measurable effects in the SPA region.

This interpretation was initially confirmed by measurements with the  $J=1\leftrightarrow2$ microwaves in the RC region switched off.  Further confirmation---and a solution to this problem---was found by deliberately tuning the $J=1\to2$ microwave frequency in the RC region around the zero-field resonance. Detuning these microwaves slightly off resonance did not affect the RC efficiency. However, it dramatically improved the $\APone$ apparent efficiency of \accumulation (Fig.~\ref{fig:SPA_efficiency_scan}\textbf{a}). 

We found that when the RC $J=1\leftrightarrow2$ microwaves were detuned $7.4~\MHz$ below the nominal $J=1 \leftrightarrow J=2$ transition frequency  ($26.669~\GHz$), the apparent \accumulation\ efficiency reached its maximum value, while  the apparent \depletion\ efficiency was unchanged. Under these conditions, apparent \depletion and \accumulation efficiencies of $\epsone^{\mathrm{\dep}} =\epsonedep$ and $\epsone^{\mathrm{\acc}} =\epsoneacc$ were achieved for \APone. Since these apparent $\APone$ transfer efficiencies are consistent with each other, we assign the actual $\APone$ transfer efficiency as the (unweighted) average of the two apparent values: $\epsone = \epsoneweighted$.

\subsection{\APtwo\ transfer efficiency}
The \APtwo\ transfer efficiency as a function of the detuning of the \APtwo\ $J=1\to2$ microwave from the expected peak value is shown in Fig.~\ref{fig:SPA_efficiency_scan}\textbf{b}. This frequency scan is performed after optimizing the \APtwo\ microwave power at the frequency of peak efficiency. With the \APtwo\ microwave detuning set for maximal \APtwo\ efficiency, the peak values of the apparent \depletion and \accumulation efficiencies are $\epstwo^{\mathrm{\dep}} = \epstwodep$ and $\epstwo^{\mathrm{\acc}} = \epstwoacc$, respectively. As before, we combine these to assign the actual \APtwo\ transfer efficiency: $\epstwo = \epstwoweighted$. Maximal \APtwo\ efficiency is achieved at $26.6684\GHz$, which is offset by $-0.38\MHz$ from the value expected from the rotational splittings and the Stark shift.

Unlike the case for \APone, the presence of the microwave field driving the $J=2\leftrightarrow3$ transition in the RC region had no measurable effect on the $\APtwo$ efficiency. However, the frequency scan in Fig.~\ref{fig:SPA_efficiency_scan}\textbf{b} shows some unanticipated features in the apparent efficiency of \accumulation (rather than a simple peak). We have qualitatively understood these features, and explain them next.

The reason for the unexpectedly low apparent efficiency of \accumulation on the low-frequency side  of the peak in {Fig.~\ref{fig:SPA_efficiency_scan}\textbf{b} can be understood from Fig.~\ref{fig:J12_transitions}. In particular: when the frequency of the microwaves driving the $J=1\to2$ transition in $\APtwo$ is shifted downward, the undesired state $\ket{J=2,F_1=3/2,F=1}$ state is Stark-shifted into resonance at a position near (or just before) the position of the \APtwo\ microwave beam's peak intensity. When this occurs, the $\APtwo$ microwaves can transfer population to this undesired level, which is not addressed by the detection laser. Hence, this effect reduces the apparent efficiency of \accumulation. 

 The diminished apparent efficiency of \accumulation over a small frequency range on the high-frequency side of the peak in Fig.~\ref{fig:J12_transitions}\textbf{b} is also understood.  This feature originates from a spatially diffuse microwave background field that arises from scattering of the (nominally Gaussian-distributed) \APtwo\ microwave field. This background field can drive population transfer from $\ket{J=1}$ into the same undesired state, and it comes into resonance with this Stark-shifted state at positions near $11~\cm$ in Fig.~\ref{fig:J12_transitions}.

\subsection{Total SPA transfer efficiency}
Combining the measured efficiencies from both $\APone$ and $\APtwo$ stages yields a total apparent efficiency (for \accumulation) of $\epstotalacc$ for the transfer $J=0\to2$ in the SPA region. 

\section{Summary and Outlook}
\label{sec:conclusion}

We have demonstrated microwave-driven adiabatic passage in a spatially varying electric field with a beam of cryogenic TlF molecules. Rotational state transfer efficiencies of $\epsone = \epsoneacc$ for \APone\ and $\epstwo = \epstwoacc$ for \APtwo\ were achieved, corresponding to a total transfer efficiency from $\ket{J=0}$ to $\ket{J=2}$ of $\epstot = \epstotalacc$ for population transfer from $J=0\to J=2$.

Since \CENTREX\ will operate in the $S=0$ state rather than the $S=1$ states studied here, we again note that the transition properties remain essentially unchanged between these two cases. For example, the transition frequencies for the $S=1$ vs.~$S=0$ cases shift by only $3\kHz$ for \APone\ and $38\kHz$ for \APtwo, far smaller than the $\gtrsim 700$~kHz range of microwave frequencies over which the state transfer efficiency $\epstot$ remains consistent with 1. The corresponding differences in Rabi rates are  $<2\pi\times1\kHz$ for \APone\ and $\approx 2\pi\times1\kHz$ for \APtwo. These changes are small compared to the current Rabi rates, and the dependence on the Rabi rates is very weak around the optimum values. Hence, we expect no noticeable change in the overall SPA state transfer efficiency when we switch to operation in the $S=0$ state.

\CENTREX\ requires two additional state-preparation regions: one to prepare the state where the Schiff moment measurement will be performed, and another to convert the final nuclear spin state into populations of different rotational states, for readout of the spin~\cite{Grasdijk_2021}. The robust and high-fidelity transfer of population into specific hyperfine states of TlF molecules that we have demonstrated here can also be applied to these other state-preparation regions in \CENTREX.  Altogether, the anticipated high efficiencies for all quantum-coherent state preparation regions will be sufficient for reaching the targeted sensitivity to the $^{205}$Tl Schiff moment in \CENTREX \cite{Grasdijk_2021}.

\begin{acknowledgments}
We are grateful for support from the John Templeton Foundation; the Heising-Simons Foundation; a NIST Precision
Measurement Grant; NSF-MRI Grants No. PHY-1827906,
No. PHY-1827964, and No. PHY-1828097; NSF Grant No.
PHY-2110420; and the Department of Energy (DOE), Office
of Science, Office of Nuclear Physics, under Contract No.
DEAC02-06CH11357 and Grant No. DE-SC0024667.
\end{acknowledgments}

\appendix

\appendix

\bibliography{references}

\end{document}